\documentclass[a4paper,reqno,twoside,psamsfonts]{amsart}
\usepackage[latin1]{inputenc}
\usepackage[T1]{fontenc}
\usepackage{textcomp}
\usepackage[british]{babel}
\usepackage{newcent} 
\usepackage[dvips,pagebackref]{hyperref}
\usepackage{bbold}
\providecommand{\href}[2]{#2}

\providecommand{\hypersetup}[1]{}
\usepackage{xspace,mathrsfs,amssymb,amsfonts}
\newcommand{\emox}[1]{%
{
\ensuremath{\mathord{#1}}
}\xspace
}
\newcommand{\CC}{{\ensuremath{\mathord{\mathbb{C}}}}\xspace}
\newcommand{\RR}{{\ensuremath{\mathord{\mathbb{R}}}}\xspace}

\newcommand{\EE}{{\ensuremath{\mathord{\mathbb{E}}}}\xspace}
\newcommand{\NN}{{\ensuremath{\mathord{\mathbb{N}}}}\xspace}

\newcommand{\ZZ}{{\ensuremath{\mathord{\mathbb{Z}}}}\xspace}
\DeclareMathAlphabet{\Bmi}{OT1}{cmm}{b}{it}

\newcommand{\HS}{{\ensuremath{\mathord{\mathcal{H}}}}\xspace} 

\newcommand{\IM}{\operatorname{Im}}

\newcommand{\Tr}{\operatorname{Tr}}

\newcommand{\wlim}{\mathop{\operatorname{w-lim}}}
\newcommand{\slim}{\mathop{\operatorname{s-lim}}}

\newcommand{\ABS}[1]{{\ensuremath{\mathord{\left|#1\right|}}}}
\newcommand{\NORM}[1]{{\mathord{\left\|#1\right\|}}}
\newcommand{\DEF}{
  \mathbin{\smash[t]{\overset{\scriptscriptstyle\mathrm{def}}{=}}}}

\newcommand{\dd}{\ensuremath{\mathrm{d}}} 
\newcommand{\ee}{\ensuremath{\mathrm{e}}} 
\newcommand{\ii}{\ensuremath{\mathrm{i}}} 
\newcommand{\ie}{, {i.e.},\xspace}
\newcommand{\eg}{, {e.g.},\xspace}

\newcommand{\GLQQ}{\char"12\kern .08em} 
\newcommand{\GRQQ}{\kern .08em\char"10\xspace}
\def\SetConstructor#1#2#3#4{%
  \def\test{#4}\ifx\test\empty{\ensuremath{\mathord{#1}_{#2}^{#3}}\xspace}%
  \else{\ensuremath{\mathord{#1}_{#2}^{#3}(#4)}\xspace}\fi}
\theoremstyle{plain}
\newtheorem{theo}{Theorem}[section]
\newtheorem*{theo*}{Theorem}
\newtheorem{defitheo}[theo]{Definition and Theorem}
\newtheorem*{defitheo*}{Definition and Theorem}
\newtheorem{prop}[theo]{Proposition}
\newtheorem*{prop*}{Proposition}
\newtheorem{coro}[theo]{Corollary}
\newtheorem*{coro*}{Corollary}
\newtheorem{lemm}[theo]{Lemma}
\newtheorem*{lemm*}{Lemma}
\theoremstyle{definition}
\newtheorem{defi}[theo]{Definition}
\newtheorem*{defi*}{Definition}
\newtheorem{defirem}[theo]{Definition and Remark}
\newtheorem*{defirem*}{Definition and Remark}
\newtheorem{rem}[theo]{Remark}
\newtheorem*{rem*}{Remark}
\theoremstyle{remark}

\newtheorem{nxmpl}[]{Example}
\newcommand{\labelS}[1]{\label{sect:#1}}
\newcommand{\labelT}[1]{\label{theo:#1}}
\newcommand{\labelP}[1]{\label{prop:#1}}
\newcommand{\labelL}[1]{\label{lemm:#1}}
\newcommand{\labelD}[1]{\label{defi:#1}}
\newcommand{\labelC}[1]{\label{coro:#1}}
\newcommand{\labelR}[1]{\label{rem:#1}}
\newcommand{\labelDR}[1]{\label{defirem:#1}}

\newcommand{\labelE}[1]{\label{eq:#1}}
\newcommand{\refC}[1]{Corollary~\ref{coro:#1}}
\newcommand{\refS}[1]{Section~\ref{sect:#1}}

\newcommand{\refP}[1]{Proposition~\ref{prop:#1}}
\newcommand{\refL}[1]{Lemma~\ref{lemm:#1}}

\newenvironment{Lprop}[2][]{\begin{prop}[#1]\labelP{#2}}{\end{prop}}
\newenvironment{Llemm}[2][]{\begin{lemm}[#1]\labelL{#2}}{\end{lemm}}

\newenvironment{Lcoro}[2][]{\begin{coro}[#1]\labelC{#2}}{\end{coro}}

\newcommand{\OA}{\emox{\mathcal{A}}}
\newcommand{\bbone}{\emox{\mathbb{1}}}
\newcommand{\XX}{\emox{\mathbb{X}}}
\begin{document}
%
%
\selectlanguage{british}
\title{Zeno Dynamics in Quantum Statistical Mechanics}
\author[A.\ U.\ Schmidt]{Andreas U.\ Schmidt}
\date{3rd January 2003, revised 5th October 2004 including corrections from the published corrigendum.}
\curraddr{Fraunhofer -- Institute Secure Information Technology\\
Dolivostraße 15\\
64293 Darmstadt\\
Germany}
\address{Fachbereich Mathematik\\
  Johann Wolfgang Goethe-Universität\\
  60054 Frankfurt am Main, Germany\\
  \href{http://www.math.uni-frankfurt.de/~aschmidt}{Homepage}}
  \email{\href{mailto:aschmidt@math.uni-frankfurt.de}{aschmidt@math.uni-frankfurt.de}}
\subjclass{\href{http://www.ams.org/msc/82Bxx.html}{82B10}; 
\href{http://www.ams.org/msc/82Cxx.html}{82C10},
\href{http://www.ams.org/msc/81Pxx.html}{81P15}}  
\thanks{\emph{PACS Subject Classification.} 
\href{http://www.aip.org/pacs/pacs01/pacs0100-02.htm}{03.65.Xp, 05.30-d, 02.30.Tb}}
\thanks{This research was supported by a research grant from the
Deutsche Forschungsgemeinschaft 
\href{http://www.dfg.de}{DFG}. The author wishes to thank the 
\href{http://www.df.unipi.it}{Dipartimento di Fisica E.~Fermi}, 
\href{http://www.unipi.it}{Universit\`{a} di Pisa}, and the
\href{http://www.infn.it}{INFN} for their hospitality.
The great work of the organisers of the conference on `Irreversible
Quantum Dynamics' at the Abdus Salam \href{http://www.ictp.trieste.it}{ICTP},
where this work has been presented, deserves special mention.
Heartfelt thanks go to  
Paolo Facchi and Saverio Pascazio (Bari, Italy),
Mark Fannes (Leuven, Belgium),
Daniel Lenz (Chemnitz, Germany),
and
Giovanni Morchio (Pisa, Italy) 
for many helpful hints and discussions.
Thanks to I.~Antoniou for pointing out some valuable references
and to the referee for calling the author's attention to some
errors in a previous version of this paper.}
\begin{abstract}
We study the quantum Zeno effect in quantum statistical 
mechanics within the operator algebraic framework.
We formulate a condition for the appearance of the
effect in 
$W^\ast$-dynamical systems, in 
terms of the short-time behaviour of the dynamics. Examples 
of quantum spin systems show that this condition can be effectively applied
to quantum statistical mechanical models. Further, we derive
an explicit form of the Zeno generator, and 
use it to construct Gibbs equilibrium states for the Zeno 
dynamics. As a concrete example, we consider the $X$-$Y$
model, for which we show that a frequent measurement at
a microscopic level\eg a single lattice site, can
produce a macroscopic effect in changing the global
equilibrium.
\end{abstract}
\maketitle
%
%
\section{Introduction}
The Zeno effect consists of an impediment to the time evolution
of a quantum system by frequent observation, for which it is
nicknamed `a watched pot never boils' or `watchdog' effect.
Research on the phenomenon has a long history dating back to the early
days of quantum theory. It found its first explicit theoretical 
formulation in~\cite{MS77}, and, after that, a vivid work in the
field was initiated, stimulated also to a great extent by significant
experimental advances. We will not review that  development here, 
but see~\cite{NNP96,WHI00} and references therein for further details. 

In~\cite{AUS02A} we followed closely the reasoning of~\cite{MS77},
and extended the theoretical treatment of the Zeno effect to modular
flows of von Neumann algebras. Our results indicate that the effect
can also appear in systems of quantum statistical mechanics at nonzero
temperature. Furthermore to a given KMS\ie equilibrium state, one will,
under favourable conditions, find an associated equilibrium state for
the Zeno dynamics\ie the limit of unitary quantum evolution interrupted
infinitely frequently by  measurement events.
This confounds the view that the induced Zeno dynamics consists 
mainly in an imposition of `boundary conditions' on the original 
dynamics~\cite{FPSS01}. To show these things, theoretically and in 
concrete examples, is our objective here.

To apply the abstract results of~\cite{AUS02A} we need a sufficient
condition for the appearance of the Zeno effect, and in particular for
the existence of the Zeno dynamics. Such a condition is derived in \refS{AZC}.
It captures the essence of the quadratic short-time behaviour of
quantum evolution~\cite{AA90,NNP96} which has since long been
identified as an essential cause for the Zeno effect. As a direct
consequence of our \textit{asymptotic Zeno condition}, we find
that the assumptions of the main Theorem~2.1 of~\cite{AUS02A}
are satisfied. Thus the Zeno dynamics will exist and form a strongly
continuous semigroup, whenever the condition holds. We derive 
this result in the context of 
$W^\ast$-dynamical systems,
to open the way for its application in quantum statistical mechanics.

The asymptotic Zeno condition is formulated in terms of the short-time
behaviour of the off-diagonal matrix elements of the original unitary
evolution with respect to the decomposition $\bbone=E+E^\perp$, where
$E$ is the projection modelling the measurement. The most pleasing 
aspect of the condition with respect to applications is that it enables
the use of perturbation theory for the examination of Zeno dynamics. 
We will show this in three examples in \refS{examples}.
The first one is a generic
example for quantum evolution impeded by the Zeno effect in
quantum statistical mechanics. 
We consider the return to equilibrium of a system which is subjected
to a bounded, local perturbation. We show that this natural relaxation 
process will be inhibited by the Zeno effect, if the measurement
controls the presence of a state which is invariant under the perturbed
dynamics.
The second example is in the general context of quantum spin systems,
and shows that the Zeno effect can decouple a finite region of the
system from its surrounding, if the interaction through the boundary
stays finite in the thermodynamic limit. 
The third example presents, as a more concrete case of the
phenomena observed in the previous two, a Zeno effect in the $X$-$Y$ 
model of an infinite spin chain.

To consider the above-mentioned question of equilibria for the Zeno dynamics, 
it is necessary to find a more explicit form of it than that provided by the limit
of infinitely frequent measurement, in which that problem would be hard to
handle. This we have already noted in~\cite[Corollary~2.2]{AUS02A}, 
where we were only able to state a formal condition for a state to be a Zeno 
equilibrium. We will gain an instrument to 
improve on that in \refS{EHE}, where we rigorously
identify the generator of the Zeno dynamics acting on the Zeno
subspace to which the dynamics becomes confined as $EHE$, where $H$ is the
original Hamiltonian. This also provides a link to the Zeno effect induced
by continuous observation in the limit of strong coupling between system
and apparatus.

Having the Zeno generator at our disposal, it is easy to construct an
important class of Zeno equilibria, namely Gibbs states, which we do
in \refS{equilibrium}. We make this explicit for quantum spin systems
and review the corresponding example~\ref{ex:qspin} of \refS{examples} 
in that respect.
The Zeno equilibrium on the bounded region in this case is the Gibbs state 
associated with a Hamiltonian which is averaged with respect to the state
imposed on the region by the given rank one projection. 
In this way, the Zeno effects implements a boundary condition on the system, 
in accordance with results of Fannes and Werner~\cite{FW95}. In the $X$-$Y$ model, 
we will be able to derive some physically remarkable results: First,
a frequent measurement on the microscopic level, even a single lattice
site, will significantly change the global equilibrium. In the concrete
example considered, it will separate the left and right subchains.
Secondly, this system will spontaneously evolve toward the Zeno
equilibrium when the observation is turned on, rendering the effect
macroscopically observable. 

Finally, the last \refS{conclusions} contains
some conclusions and an outlook to possible further 
applications in physical models.

It should be noted that we 
restrict our discussion completely to a concrete realisation
of a 
$W^\ast$-dynamical system given by the GNS
representation $\pi_\omega$ of a fixed, \textit{a priori} chosen 
KMS state $\omega$. That is we consider the 
von Neumann algebra 
$\pi_\omega(\OA)$ on the GNS Hilbert space \HS and assume the 
dynamical automorphism group to be $\pi_\omega$-covariant\ie to 
be realised by a strongly continuous, unitary group of operators. 
This notably simplifies our treatment, but also restricts it
to a single superselection sector of the theory. Nevertheless,
the results in Sections~\ref{sect:EHE} and~\ref{sect:equilibrium}
regarding Zeno equilibria are essentially
independent of the chosen representation.
\section{A Sufficient Asymptotic Condition for Zeno Dynamics}
\labelS{AZC}
The Zeno effect is commonly attributed to the quadratic short-time
behaviour of quantum evolution~\cite{PL98} which in turn is rooted
deeply in the geometry of Hilbert space~\cite{AA90}.
This quadratic behaviour seems so generic that one can hope
to turn it into a sufficient condition for the effect to occur.
This is what we will present in this section.

Let $E$ be a projection, $U$ a unitary group on a Hilbert space \HS, and set
\[
F_n(t)\DEF \bigl[E U(t/n) E\bigr]^n, \quad
\text{for }t\in\RR,\ n\in\NN.
\]
This is a time evolution of a system interrupted by frequent,
instantaneous `measurement' effects, coarsely modelled by projections
(or, if one wishes, employing the projection postulate). The question
whether the strong Zeno effect, or `Zeno paradox' occurs is in essence
equivalent to the question of strong convergence of the operator
sequence $F_n$ to a sensible\ie continuous, time 
evolution~\cite{MS77}. For then, the induced evolution will be confined to a
`Zeno subspace' within $E\HS$ by `infinitely frequent observation.' This limit
is arguably unphysical~\cite{PAT96}, but still of conceptual interest, as 
will become evident below.

To exemplify the basic idea of our proceeding, we want
to see whether the $F_n(t)$ form a Cauchy sequence in $n$ for given $t$.
For that, we have to estimate the quantities
\[
  \NORM{\bigl(F_n(t)-F_m(t)\bigr)}\leq
  \NORM{\bigl(F_n(t)-F_{nm}(t)\bigr)} + 
  \NORM{\bigl(F_m(t)-F_{nm}(t)\bigr)}.
\]
A double telescopic estimation yields
\begin{multline*}
  \NORM{\bigl(F_n(t)-F_{nm}(t)\bigr)} \leq\\
  \sum_{k=1}^{n}\sum_{l=1}^{m-1}
  \left\|
  \bigl[EU(t/n)E\bigr]^{n-k} 
  \left( EU(t(m-l)/(nm))E \bigl[EU(t/(nm))E\bigr]^{l} -\right.\right.\\
   \left.\left. EU(t(m-l+1)/(nm))E \bigl[EU(t/(nm))E\bigr]^{l-1} \right)
   \bigl[EU(t/(nm))E\bigr]^{m(k-1)}\right\| .
\end{multline*}
Now, since with $E^{\perp}\DEF\bbone-E$ we have
\[
EU(t(m-l+1)/(nm))E = E U(t (m-l)/nm)(E+E^\perp )  U(t/(nm)) E, 
\]
we find that the $(k,l)$th term in the sum is equal to
\begin{multline*}
  \left\|
  \bigl[EU(t/n)E\bigr]^{n-k} 
  \cdot EU(t(m-l)/(nm)) E^\perp \cdot \right.\\
  \left. \cdot E^\perp U(t/(nm)) E \cdot \bigl[EU(t/(nm))E\bigr]^{l-1}
   \bigl[EU(t/(nm))E\bigr]^{m(k-1)}\right\| .
\end{multline*}
Multiplying out and using repeatedly $\NORM{AB}\leq\NORM{A}\NORM{B}$, we estimate this expression
from above by
\begin{multline*}
  \NORM{\bigl[EU(t/n)E\bigr]}^{n-k} 
  \cdot \NORM{EU(t(m-l)/(nm)) E^\perp}\\
  \cdot \NORM{E^\perp U(t/(nm)) E }\cdot \NORM{\bigl[EU(t/(nm))E\bigr]}^{l-1}
   \NORM{\bigl[EU(t/(nm))E\bigr]}^{m(k-1)} .
\end{multline*}
Observing that all terms containing only the projection $E$ have operator norm $\leq1$
and can thus be omitted in the estimation of $\NORM{\bigl(F_n(t)-F_{nm}(t)\bigr)}$,
we arrive at
\[
  \NORM{\bigl(F_n(t)-F_{nm}(t)\bigr)} \leq  
   \sum_{k=1}^{n}\sum_{l=1}^{m-1}
  \NORM{E U(t(m-l)/(nm)) E^\perp}\NORM{ E^\perp U(t/(nm)) E }.
\]
Now, we require
$E^\perp U(\tau)E=O(\tau)$ uniformly as $\tau\to0$. 
That is,
there shall exist $\tau_0>0$ and $C\geq0$ such that for all $\tau$ with 
$\ABS{\tau}<\tau_0$ holds
the estimate
$\NORM{E^\perp U(\tau)E}\leq C^{1/2}\ABS{\tau}$. 
Then, for $n>n_0\geq1/\tau_0$, and $m\geq2$,
\begin{align*}
  \NORM{\bigl(F_n(t)-F_{nm}(t)\bigr)} &\leq
  C
  t^2\sum_{k=1}^{n}\sum_{l=1}^{m-1}
  \frac{m-l}{n^2m^2}\\& =
  C
t^2\sum_{k=1}^{n} \frac{(m-1)m}{2n^2m^2}\\ & =
  \frac{C
t^2}{2} \frac{(m-1)m}{n m^2}\leq 
  \frac{C
t^2}{2n}.
\end{align*}
An analogous estimate holds for
$\NORM{\bigl(F_m(t)-F_{nm}(t)\bigr)
}$, which yields
for $m-2\geq n>n_0\geq1/\tau_0 $ the overall result
\begin{equation}\labelE{cauchy-est}\tag{$\ast$}
 \NORM{\bigl(F_n(t)-F_m(t)\bigr)
}\leq
 \frac{C
t^2}{n}.
\end{equation}
We have proved the essence of
\begin{Llemm}{Ot-sufficient}
  If $E^\perp U(\tau)E=O(\tau)$ uniformly as $\tau\to0$ then
  $F_n(t)$ converges 
  uniformly as $n\to\infty$ for all $t\in\RR$.
  Furthermore $W(t)\DEF\slim_{n\to\infty}F_n(t)$ is 
  uniformly
  continuous in $t$ and $\slim_{t\to0}W(t)=E$.
\end{Llemm}
\begin{proof}
  The first statement is clear since \eqref{eq:cauchy-est} shows
that the $F_n(t)$ form Cauchy sequences which are therefore
\textit{a fortiori} convergent. The other statements follow
from from $F_n(0)=E$ for all $n$, and the fact that the convergence 
of $F_n(t)$ is uniform for $t$
on compact subsets of \RR. This follows in turn from the
$t$-dependence of the estimate~\eqref{eq:cauchy-est}.
\end{proof}
We will now use the above result to reformulate the main Theorem~2.1 of~\cite{AUS02A}
in a more effective way. The general setting is as follows:
Let $(\OA,\tau)$ be a 
$W^\ast$-dynamical
system with faithful $(\tau,\beta)$-KMS state $\omega$ 
which is assumed to be normal.
Denote by $\Omega$ the vector representative
of $\omega$ in the associated representation $\pi_\omega$ on the GNS--Hilbert space
$\HS$.  
The automorphism group $\tau$ is assumed to be implemented covariantly\ie
by a strongly continuous group of unitary operators $U(t)$
on $\HS$. The representation
$\pi_\omega$ will be omitted from the notation, 
when no confusion is possible. 
\begin{Lprop}{ZDexist}
Under the conditions described above, let $\beta>0$, assume $\OA$ to be unital, 
let $E\in\OA$ be a projection, and set $E^\perp\DEF\bbone-E$. Assume that
the \textbf{asymptotic Zeno condition} holds:
For $A\in\OA$, the estimate
\begin{equation}
\labelE{AZC}\tag{AZC}
\NORM{E^\perp U(\zeta) E A\Omega} \leq C\cdot\NORM{A\Omega}\cdot \ABS{\zeta}
\end{equation}
is valid for $\zeta$ with $\ABS{\zeta}<r_0$ for some fixed 
$r_0>0$ and $\IM\zeta\geq 0$. 
In short: $(U,E)$ satisfies~\eqref{eq:AZC} for $\OA$.
Then the strong operator limits
\[
W(t)\DEF\slim_{n\to\infty}
\bigl[E U(t/n) E\bigr]^n
\]
exist, and form a strongly continuous group of unitary operators
on the \textbf{Zeno subspace} $\HS_E\DEF\overline{\OA_E\Omega}\subset E\HS$,
where $\OA_E\DEF E\OA E$. The group $W(t)$
induces an automorphism group $\tau^E$ of  
$\OA_E$, 
such that $(\OA_E,\tau^E)$ is a 
$W^\ast$-dynamical system. The vectors 
$W(z)A_E\Omega$,  $A_E\in\OA_E$, extend analytically 
to the strip $0<\IM{z}<\beta/2$ and 
are continuous on its boundary.
\end{Lprop}
\begin{proof}
We show that the assumptions 
of the main Theorem~2.1 of~\cite{AUS02A} are satisfied, 
from which we obtain the stated conclusions.
First, for real $\tau$, \eqref{eq:AZC} implies $E^\perp U(\tau) E=O(\tau)$ 
uniformly since the operators in question are bounded, $\OA\Omega$ is dense in $\HS$,
and~\eqref{eq:AZC} is uniform in $A$ on a fixed real neighbourhood of $0$. 
Therefore \refL{Ot-sufficient} yields the existence
of $W(t)$, $t\in\RR$, its weak continuity in $t$ and the initial condition
$\wlim_{t\to0}W(t)=E$. These facts comprise condition~i)
of~\cite[Theorem~2.1]{AUS02A} (keeping in mind, here and in the following,
the connection between faithful states of von Neumann algebras and KMS states
given by Takesaki's theorem~\cite[Theorem~5.3.10]{b:BR79/81}).
For the second condition of the cited theorem, we need only to show
that $W(t+\ii\beta/2)$ exist as strong operator limits on the common, dense
domain $\OA\Omega$.
For this notice that the calculations leading
to~\eqref{eq:cauchy-est} are applicable to  
$(\smash{F_n(t+\ii\beta/2)-F_m(t+\ii\beta/2)})A\Omega$, 
leading to the estimate
\[
 \NORM{\bigl(F_n(t+\ii\beta/2)-F_m(t+\ii\beta/2)\bigr)
A\Omega
}\leq
 \frac{C
 \NORM{A}
\ABS{t+\ii\beta/2}^2}{n}
\]
for $A\in\OA\Omega$, and $m-2\geq n> n_0\geq 1/r_0$.
Thus, also condition~ii) of Theorem~2.1 of~\cite{AUS02A} is
satisfied and the stated conclusions follow from it.
\end{proof}
Note that it would have been sufficient to test the asymptotic 
condition on any dense set of
vectors which are analytic for $U(\zeta)$, 
in some strip $0\leq\IM\zeta<\varepsilon$ for some 
$\varepsilon\geq\beta/2$.
For simplicity, we restricted attention to $\OA\Omega$.
The~AZC is strictly stronger than the assumptions
of \cite[Theorem~2.1]{AUS02A}, where no continuity
at the boundary $\IM z = \beta/2$ was implied
(as would also follow from the estimate in the proof above)
and only weak continuity at the real axis needed to be assumed.

In the examples below weak closures will be understood
for all observable algebras\ie we will always consider
the von Neumann algebra $\pi_\omega(\OA)''$. 
That is, we will not exert the discrimination
between $C^\ast$- and $W^\ast$-dynamical systems, which is
not very important for our purpose here. The main merit of this 
simplification is that $\tau^E$ is guaranteed by~\cite[Lemma~2.6]{AUS02A}
to act by automorphisms on the von Neumann subalgebra $\OA_E$ of $\OA$.
However it seems possible to obtain a result corresponding 
to~\refP{ZDexist} in the $C^\ast$ case, repeating the arguments
of~\cite{AUS02A} using the analogous analyticity properties
of the vectors in $\OA\Omega$ as detailed 
in the proof of~\cite[Theorem~5.4.4]{b:BR79/81}.

The condition~\eqref{eq:AZC} is quite weak and thus 
indicates how generic a quantum phenomenon the Zeno effect indeed is.
For example it is always satisfied if the generator $H$ of the group $U$
is bounded, or, more generally, if $E$ projects onto
a closed subspace of entire analytic elements for $H$\eg
if $E$ is contained in a bounded spectral projection of $H$. 
In those cases a power series expansion of $U(t)=\ee^{\ii t H}$
implies~\eqref{eq:AZC}.
However, if neither is the case, then~\eqref{eq:AZC} will generally 
fail in that its defining estimate is not uniform in $A\Omega\in\HS$.

It is also noteworthy that in showing the convergence of $F_n$ to $W$,
we have not used the unitarity of $U$. Thus an analogue of the Zeno
effect is also possible for non-unitary (non-Hamiltonian, 
non-Schrödinger) evolutions, cf.~\cite{PL98}. 
On the other hand, the group property of $U$ was essential for 
obtaining the quadratic term that forced
the convergence of the sequence. 

Condition~\eqref{eq:AZC} 
is comparable to other conditions for the appearance of the 
Zeno effect, which are commonly based on the finiteness of the moments 
of the Hamiltonian in the Zeno subspace~\cite{NNP96}.
The AZC, when restricted to the real axis,
is equivalent to saying that the function $E^\perp U(t) E$ is
uniformly Lipschitz continuous at the point $t=0$. Not surprisingly,
Lipschitz continuity is well known as a salient condition for the 
existence of solutions to (nonlinear) evolution equations.

Asymptotic bounds on $E^\perp U(t) E$ have already been considered 
by other authors~\cite{EX89, NIS88, MS77A}, in the context of short-time
regeneration of an undecayed state. In particular in~\cite{EX89}, the
deviation of the `reduced evolution' $EU(t)E$ from being a semigroup has 
been expressed by such (polynomial) bounds. We will obtain a similar
yet somewhat coarser result in \refS{EHE}.
\section{Examples}\labelS{examples}
The power of~\eqref{eq:AZC} lies to a great extent in that it
yields perturbative conditions for the occurrence of the Zeno effect.
For it is known that a perturbed semigroup $U^P_t$, resulting from adding
a bounded perturbation $P$ to a $C_0$-semigroup $U_t$, is close to $U_t$
for small times in the sense that
$\NORM{U_t-\smash{U^P_t}}=O(t)$, as $t\to0$, 
see~\cite[Theorem~3.1.33]{b:BR79/81}. Now if
$E$ projects onto a subspace which is invariant under $U_t$, then
this asymptotic behaviour implies that the Zeno dynamics of the 
pair $(U^P_t,E)$ exists. We exemplify this basic mechanism in
the following.
\begin{nxmpl}[Non-Return to Equilibrium]
  It is well known~\cite{ROB73} that a quantum system will under
general conditions\eg if $(\OA,\tau)$ is asymptotically Abelian, 
return to equilibrium for large times. 
This means the following:  
If the system is prepared in an equilibrium state $\omega^P$
for the perturbed evolution $\tau^P$, 
where $P=P^\ast\in\OA_\tau$ is a bounded perturbation, which is
in the set of entire analytic elements $\OA_\tau$ 
for $\tau$ (termed local perturbation), 
and thereafter evolves under the unperturbed 
dynamics $\tau$, one recovers a $\tau$-equilibrium state $\omega_±$ 
for $t\to±\infty$. 
Assume that the perturbed and unperturbed dynamics
are implemented by unitaries $U^P$ and $U$, respectively. 
This is always possible if either $\tau$ or $\tau^P$ 
is covariant in the chosen representation~\cite[Theorem~1]{ROB73}.
Then, the unperturbed dynamics can be written
in terms of the perturbed one by the perturbation
expansion~\cite[Theorem~3.1.33 and Proposition~5.4.1]{b:BR79/81}
\[
      U(t)=U^P(t)+ 
      \sum\limits_{n\geq1}\int\limits_0^t\dd t_1\dotsi\int\limits_0^{t_{n-1}}\dd t_n
      U^P(t_1) P U^P(t_2-t_1)P \cdots P U^P(t-t_n), 
\]
where the $n$-th term in the sum is bounded by $\NORM{P}^nt^n/n!$.
Let the system be prepared
in any $\tau^P$-invariant state $\varphi^P$. 
In the representation $\pi^P$
induced by the chosen $\tau^P$-KMS state $\omega^P$
the corresponding vector states are denoted by
$\Phi^P$ and $\Omega^P$ respectively. Let $E$ be the projection onto
the space spanned by the vector $\Phi^P$ and assume $E\in\OA$.  Then the above expansion
readily yields $E^\perp U(t) E=O(t)$ uniformly,
since the $\tau^P$-invariance of $\varphi^P$ 
implies $U^P(t)\Phi^P=\Phi^P$. In application to vectors in
$\OA\Omega$ this estimate extends to a fixed, small neighbourhood
of $0$ in the upper half-plane and is uniform in those vectors.
Thus~\eqref{eq:AZC} holds, the Zeno dynamics converges, and
the system remains in the state $\varphi^P$. The same reasoning
is applicable if $E$ projects onto a $\tau^P$-invariant subspace.
\end{nxmpl}
The phenomenon described in this example is the direct counterpart,
in the context of quantum statistical mechanics,
of the most common example for the Zeno effect in quantum mechanics\ie 
the prevention of a decay process, see\eg~\cite{HNNPR98,FGMPS00}.
Its character is generic, and therefore we formulate it as a corollary.
\begin{Lcoro}{non-return}
  Let $(\tau,\OA)$ be as above. 
Let $P\in\OA$ be a local perturbation,
and denote by $\tau^P$
perturbed dynamics as constructed 
in~\cite[Proposition~5.4.1 and Corollary~5.4.2]{b:BR79/81}.
Let $E\in\OA$ be a $\tau^P$-invariant projection\ie $\tau^P(E)=E$. 
Then the $(\tau,E)$-Zeno dynamics
$\tau^E$ 
is an automorphism group of 
$\OA_E$, and $\HS_E$ is $\tau^{E}$-in­va­ri­ant.
\end{Lcoro}
 In view of the mechanism noted at the end of Example~\ref{ex:qspin} below, 
this corollary could easily be reformulated in terms of the 
thermodynamic limit of local algebras over bounded regions. 
We omit the details.
\begin{nxmpl}[%
\renewcommand{\thefootnote}{\ensuremath{\fnsymbol{footnote}}}%
Local Domains of Quantum Spin Systems\footnote{This example
was suggested by G.~Morchio.}]
\label{ex:qspin}
For a detailed exposition of the notions and facts invoked below,
we refer the reader to~\cite[Section~6.2]{b:BR79/81}.
Consider a quantum spin system over the lattice $\XX\DEF\ZZ^d$ with 
interaction $\Phi\colon \XX\supset X\longmapsto \OA_X$. 
The local Hamiltonian of a bounded subset $\Lambda\subset\XX$ is
$H_\Phi(\Lambda)\DEF\sum_{X\subset \Lambda}\Phi(X)$ and
$U_{\Lambda}(t)\DEF\ee^{\ii tH_\Phi(\Lambda)}$ 
is the associated group of unitaries.
Consider bounded
subsets $\Lambda\subset\Lambda'\subset\XX$.
The surface interaction of $\Lambda$ with $\Lambda'$ is
\[
W_\Phi(\Lambda;\Lambda')\DEF\sum\bigl\{
\Phi(X)\bigm| X\subset\Lambda',\ 
X\cap\Lambda'\setminus\Lambda\neq\varnothing,\
X\cap\Lambda\neq\varnothing
\bigr\}.
\]
Then holds the decomposition
\[
H_\Phi(\Lambda')=H_\Phi(\Lambda'\setminus\Lambda)+
H_\Phi(\Lambda)+W_\Phi(\Lambda;\Lambda'),
\]
and $\bigl[ H_\Phi(\Lambda'\setminus\Lambda), H_\Phi(\Lambda) \bigr]=0$.
Let $\Phi_{\Lambda}$ be any vector in $\HS_\Lambda$ and
$\varphi_{\Lambda}$ the
associated local pure state 
of the closed subsystem localised in $\Lambda$. 
Define a projector
on $\HS_{\Lambda'}=\HS_{\Lambda'\setminus\Lambda}\otimes\HS_\Lambda$ by
\[
E_{\varphi_\Lambda;\Lambda'}
\DEF \bbone_{\Lambda'\setminus\Lambda}\otimes P_{\Phi_{\Lambda}},
\]
where $P_{\Phi_{\Lambda}}$ 
is the projector onto the one-dimensional
subspace generated by $\Phi_{\Lambda}$ 
in $\HS_{\Lambda}$ (note that this rank one projector is always
in $\OA_\Lambda=\mathcal{B}(\HS_\Lambda)$ for bounded regions $\Lambda$).
Now $P_\Phi(\Lambda;\Lambda')\DEF H_{\Phi}(\Lambda)+W_\Phi(\Lambda; \Lambda')$ is the
local perturbation which removes the effect of the region $\Lambda$
on the local system in $\Lambda'$. The state $\varphi_\Lambda$ is clearly invariant
under the perturbed dynamics generated by 
$H_\Phi(\Lambda')-P_\Phi(\Lambda;\Lambda')$
and therefore \refC{non-return} can be applied.
Thus the local limit dynamics
\[
W_{\varphi_\Lambda;\Lambda'}(t)=\lim_{n\to\infty} 
\bigl[E_{\varphi_\Lambda;\Lambda'}U_{\Lambda'}(t/n) 
E_{\varphi_\Lambda;\Lambda'}\bigr]^n
\]
is well defined. This result persists in the thermodynamic limit
if the global interaction energy
\[
W_\Phi(\Lambda)\DEF
\sum\bigl\{\Phi(X)
\bigm|X\cap\Lambda\neq\varnothing,\ X\cap\Lambda^c\neq\varnothing\bigr\}
=\lim_{\Lambda'\to\infty}W_\Phi(\Lambda;\Lambda')
\]
is well defined, and then the local limits are uniform in $\Lambda'$.
Choose a faithful normal state for the global algebra $\OA$, and
let $U$ be the unitary generating the global dynamics $\tau$ in its
GNS representation.
Under the assumptions above we obtain that the Zeno limit
\[
W_{\varphi_\Lambda}(t)=
\lim_{n\to\infty} 
\bigl[E_{\varphi_\Lambda}U(t/n) E_{\varphi_\Lambda}\bigr]^n
\]
exists, where $E_{\varphi_\Lambda}\DEF\lim_{\Lambda'\to\infty}E_{\varphi_\Lambda}=
\bbone_{\Lambda^c}\otimes P_{\Phi_\Lambda}$.
It defines an automorphism group $\tau^E$ of $\OA$.
Furthermore, this automorphism group is the uniform limit of the
local automorphism groups  $\tau^{E_{\varphi_\Lambda;\Lambda'}}$ defined by $W_{\varphi_\Lambda;\Lambda'}$.
\end{nxmpl}
This simple method to obtain Zeno dynamics will in general only work
for projections over bounded regions. The requirements one would have
to fulfil over unbounded regions are severe: If we want to apply 
the perturbative method, the Hamiltonian over that region would
have to be bounded. Or, if one looks for projectors onto 
states over the unbounded region which are invariant from the outset, 
those might be scarce
if the system is sufficiently disordered\eg asymptotically Abelian.
As a negative example, the projector onto a KMS equilibrium state
over an unbounded region is generically not an observable.
\begin{nxmpl}[The $X$-$Y$ Model]
\label{ex:XY}
We want to illustrate the two facets of the Zeno effect exhibited
above in the more concrete model of the $X$-$Y$-spin chain.
This model has been rigorously 
treated in~\cite{ROB73}, where all the facts used below are proved.
It consists of a spin chain over $\ZZ$, where the state space over
a point $x\in\ZZ$ is two-dimensional $\HS_x\DEF\CC^2$. The local
algebras over a bounded region $\OA_{[n,m]}$, $n\leq m\in\ZZ$, 
are generated by the fermionic generation and annihilation operators
$a_x$, $a^\ast_x$, $n\leq x\leq m$, with commutation relations
$[a_x,a_y]=0=[a_x,a_y^\ast]$, $x\neq y$, and $\{a_x,a_x^\ast\}=1$,
$\{a_x,a_x\}=0$, where $\{\cdot,\cdot\}$ denotes the anti-commutator.
The global algebra \OA is the weak closure of the union of
the $\OA_{[-n,n]}$.
The local dynamics is given by the Hamiltonian
\[
H_{[n,m]}\DEF \frac{J}{2}
\sum_{x=n}^{m-1} \bigl(a_x^\ast a_{x+1} +  a_{x+1}^\ast a_x \bigr) +
h \sum_{x=n}^m a_x^\ast a_x.
\]
The global dynamics in the thermodynamic limit
\[
\tau_t(A)\DEF\lim_{n\to\infty}
\ee^{\ii t H_{[-n,n]}} A_n \ee^{-\ii t H_{[-n,n]}},
\] 
for $A=\lim_{n\to\infty}A_n\in\OA$, $A_n\in\OA_{[-n,n]}$, $t\in\RR$,
exists, and renders $(\OA,\tau)$ a $W^\ast$-dynamical system, whose
unique $(\tau,\beta)$-KMS state at given inverse temperature $\beta$ is the
unique weak-* limit $\omega_\beta$ of any increasing net of local Gibbs states over $[n,m]$.
Now let $P_0$ be the perturbation which removes the particle at
position $0$:
\[
P_0\DEF - \frac{J}{2} 
\bigl( a_{-1}^\ast a_{0} + a_{0}^\ast a_{1} 
+  a_{0}^\ast a_{-1} + a_{1}^\ast a_{0} \bigr)
- h a_0^\ast a_0,
\]
such that the Hamiltonian over $[-n,n]$, $n\geq 1$, decomposes as
\[
H_{[-n,n]}=H_{[-n,-1]}+P_0+H_{[1,n]}.
\]
Let $\omega_{L,\beta}$, $\omega_{R,\beta}$ be the Gibbs equilibrium
states over the subchains $[-\infty,-1]$ and $[1,\infty]$, respectively,
obtained as limits of local Gibbs states at inverse temperature $\beta$, 
and let $\rho_0$ be an arbitrary
state over $\OA_0$. Then the product state
\[
\varphi_{\rho_0,\beta}\DEF\omega_{L,\beta}\otimes\rho_0\otimes\omega_{R,\beta}
\]
is invariant under the perturbed dynamics $\tau^{P_0}$. But as 
$(\OA,\tau)$ is asymptotically Abelian, it follows that return to
equilibrium will occur\ie
\[
\lim_{\ABS{t}\to\infty}\varphi_{\rho_0,\beta}(\tau_t(A))=\omega_\beta(A),
\quad\text{for }A\in\OA.
\]
If we choose $E$ to be the projection onto a vector
representative of $\rho_0$ at site $0$ and the identity on
the rest of the chain, then we have a special instance of \refC{non-return}. 
Thus, the Zeno dynamics $\tau^E$ exists and prevents the return
to the global equilibrium. The left and right subchains remain dynamically
isolated, and the arbitrary state $\rho_0$ at the point $0$ is preserved.
The state $\varphi_{\rho_0,\beta}$ is preserved by the perturbed
evolution and by $E$, however it is not a good candidate
for a genuine equilibrium state for the Zeno dynamics $\tau^E$. We will
return to that matter at the end of \refS{equilibrium}
\end{nxmpl}
\section{The Explicit Form of the Zeno Hamiltonian}
\labelS{EHE}
We want to show that if the Zeno dynamics converges, 
it is possible to identify its generator explicitly. This will
become useful in the sequel.

Let $H$ be the generator of $U(t)=\ee^{\ii tH}$. 
The unitary group $U_E(t)\DEF\ee^{\ii t EHE}$ is 
called the \textbf{reduced dynamics} associated to $(U,E)$.
Notice that $U_E$ induces an automorphism group $\widehat{\tau}^E$
of $\OA_E$ whenever the group $\tau$  is one for \OA.
To be able to compare the reduced with the Zeno dynamics, we need
a technical condition:
We call $(U,E)$ \textbf{regular} if $\OA_E$ contains a dense
set of elements which are analytic for $\tau$ in an arbitrary 
neighbourhood of zero. The condition of regularity will be
required to have enough analytic vectors in $\HS_E$ at hand 
for the proof below to work. It excludes pathological cases\eg 
when $E$ projects onto a subspace of states with properly infinite energy. 
It is automatically satisfied in all examples we consider, see
the comment after the proof of the following proposition.
\begin{Lprop}{reduced}
  Let $(U,E)$ be regular and satisfy~\eqref{eq:AZC} for \OA. 
  Then $U_E(t)$ equals $W(t)$, when restricted to $\HS_E$.
\end{Lprop}
  Throughout the proof below let $\Psi_E\in \OA_{E,\tau}\Omega\subset\HS_E$, where
$\OA_{E,\tau}$ is a dense set of  elements in $\OA_E$, which are analytic for $\tau$.
Record that, by the discussion following~\cite[Definition~3.1.17]{b:BR79/81}, 
the $\tau$-analyticity
of $\Psi_E$ is equivalent to analyticity with respect 
to $U$ and this is in turn equivalent to the convergence of power series
of analytic functions in $\sigma H$ applied to $\Psi_E$, 
for $\sigma\in\CC$ small enough,
as given in the cited definition.
\begin{proof}[{Proof of~\refP{reduced}}]
We first derive a useful asymptotic estimate:
Setting $\Psi_E(\sigma)\DEF U_E(\sigma)\Psi_E$ holds 
\begin{align*}
  \NORM{\bigl(U_E(\tau) - EU(\tau)E\bigr)\Psi_E(\sigma)} &= 
  \NORM{ \left\{ \sum_{k=0}^\infty \frac{(\ii\tau)^k (EHE)^k}{k!}   - 
        E \sum_{l=0}^\infty \frac{(\ii\tau)^l H^l}{l!} E \right\} \Psi_E(\sigma)} \\
  &= \NORM{\sum_{k=2}^\infty\frac{(\ii \tau)^k}{k!} \bigl[ (EHE)^k  - EH^kE \bigr]\Psi_E(\sigma)},
\intertext{using $E\Psi_E(\sigma)=\Psi_E(\sigma)$, 
which is clear since $U_E$ commutes with $E$. 
By using $\NORM{E}=1$, this can be estimated further as}
  &\leq
  2 \sum_{k=2}^\infty\frac{\ABS{\tau}^k}{k!}\NORM{H^k\Psi_E(\sigma)}.\\
\end{align*}
Since $\Psi_E$ is analytic for $U$ in a neighbourhood of $0$, 
also the translates $\Psi_E(\sigma)=U_E(\sigma)\Psi_E$, 
for $\sigma$ small enough, will be analytic for $U$ in a somewhat smaller
neighbourhood of $0$. This can be seen by noting that the power series
of $U_E(\sigma)$ is term-wise bounded in norm by a convergent one, where
$EHE$ is replaced by $H$, using $\NORM{E}=1$. The composition
of power series in question then amounts to the composition
of analytic functions of $H$ for $\sigma$, $\tau$, small enough.
Therefore the power series on the right hand side of the last inequality
is convergent for $\sigma$, $\tau$ small, 
and defines an analytic function in $\tau$ which is $O(\ABS{\tau}^2)$ as 
$\ABS{\tau}\to0$.
Thus, we finally obtain for small enough $\sigma$, $\tau$ the estimate
\begin{equation}
  \label{eq:UE-EUE-est}
  \tag{$\dag$}
  \NORM{\bigl(U_E(\tau) - EU(\tau)E\bigr)\Psi_E(\sigma)}\leq
  \tau^2\cdot C_{\Psi_E,\sigma}<\infty.
\end{equation}
Now, from $U_E(t)\Psi_E=EU_E(t)E\Psi_E$, follows the identity
\begin{equation}
  \label{eq:UE-insert}
  \tag{$\ast\ast$}
  U_E(t)\Psi_E=\bigl[EU_E(t/n)E\bigr]^n\Psi_E, \quad\text{for all } n,
\end{equation}
by iteration.
Exploiting this, we can rewrite $F_n(t)-U_E(t)$ to yield
\begin{align*}
  \bigl\| & F_n(t)\Psi_E - U_E(t)\Psi_E\bigr\| =
  \NORM{\bigl[ EU(t/n)E \bigr]^n \Psi_E - \bigl[ EU_E(t/n) E\bigr]^n \Psi_E}.\\
\intertext{A telescopic estimate shows}
 & \leq 
 \sum_{i=1}^n \NORM{\left\{ 
   \bigl[ EU(t/n)E \bigr]^{n-i}
   \bigl( EU(t/n)E - EU_E(t/n)E \bigr)
   \bigl[ EU_E(t/n)E \bigr]^{i-1}
 \right\}\Psi_E}.
\end{align*}
The norm of the vector under the sum is, using~\eqref{eq:UE-insert},
\begin{align*}
  & \NORM{ 
   [ EU(t/n)E ]^{n-i}
   ( EU(t/n)E - EU_E(t/n)E )
   \Psi_E(t(i-1)/n)}\\
\intertext{Using commutativity of $U_E$ with $E$, 
and the invariance of $\Psi_E(\sigma)$ under $E$, we have
$EU_E(t/n)E  \Psi_E(t(i-1)/n)=U_E(t/n) \Psi_E(t(i-1)/n)$,
and use this to rewrite the above expression as}
= & \NORM{ [ EU(t/n)E ]^{n-i} ( EU(t/n)E - U_E(t/n) ) \Psi_E(t(i-1)/n)}\\
\intertext{Now, with $\NORM{\smash{ [ EU(t/n)E ]^{n-i}} }\leq1$ and 
$\NORM{AB\Psi}\leq\NORM{A}\NORM{B\Psi}$, this is bounded by}
\leq &  \NORM{ ( EU(t/n)E - U_E(t/n) ) \Psi_E(t(i-1)/n)}.
\end{align*}
Putting this together, we obtain the estimate
\[
\bigl\|  F_n(t)\Psi_E - U_E(t)\Psi_E\bigr\| \leq
\sum_{i=1}^n \NORM{ ( EU(t/n)E - U_E(t/n) ) \Psi_E(t(i-1)/n)}.
\]
We can now apply~\eqref{eq:UE-EUE-est} to obtain,
for $n>M$ large enough,
\[
 \NORM{ F_n(t)\Psi_E - U_E(t)\Psi_E} \leq \sum_{i=1}^n 
 \left(\frac{t}{n}\right)^2 \cdot 
 \sup_{\ABS{\sigma}\leq\ABS{t}}C_{\Psi_E,\sigma}=
 \frac{t^2C'_{\Psi_E,t}}{n},
\]
for some finite $C'_{\Psi_E,t}$.
Since $F_n$ converges strongly to $W$ by~\eqref{eq:AZC}, it
follows $W(t)\Psi_E=U_E(t)\Psi_E$.
The density of the elements $\OA_{E,\tau}\Omega$ 
in $\HS_E$ then shows the claim.
\end{proof}
Let us return to Example~\ref{ex:qspin} to see that the regularity
condition is satisfied there. If $\Psi$ is an entire analytic vector
for $U$, then applying the projector $E_{\varphi_\Lambda}$ changes
its total energy only by a finite amount, since the perturbation
$P_\Phi(\Lambda)\DEF H_{\Phi}(\Lambda)+W_\Phi(\Lambda)$
is bounded. A finite change of energy does not affect 
the analyticity with respect to $\tau$ and thus the projected
vector is $\tau$-analytic.
Since one always finds a dense set of analytic elements
in $\OA$~\cite[Proposition~2.5.22]{b:BR79/81}, the regularity 
condition is satisfied in this case. This argument is also valid
under the general assumptions of  \refC{non-return}.

The explicit form of the generator for the Zeno dynamics
also yields an heuristic argument for the equivalence of
the Zeno effects produced by `pulsed' and `continuous'
measurement, respectively. The latter commonly denotes
the simple model for the coupling of the quantum system to a
measurement apparatus by adding to the original Hamiltonian a measurement Hamiltonian 
multiplied by a coupling constant, and letting
the coupling constant tend to infinity~\cite{FP01,FP02,FP02B}. 
The essential point here is that the degrees of freedom
in the Zeno subspace $\HS_E$ become energetically infinitely
separated from those in its orthogonal complement. For
this it suffices to set
\[
H_K\DEF H + K E^\perp,\quad U_K(t)\DEF\ee^{\ii t H_K},
\]
and to consider the limit $K\to\infty$. This can be 
done by applying analytic perturbation theory to
\[
H_\lambda\DEF\lambda H+E^\perp, \quad \text{with }\lambda\DEF K^{-1},
\]
and
\[
U_\lambda(\tau)\DEF\ee^{\ii \tau H_\lambda}=U_K(t), 
\quad \text{with } \tau\DEF Kt=t/\lambda,
\]
around $\lambda=0$. The final result is
\[
\lim_{K\to\infty}U_K(t)\Phi=\ee^{\ii t EHE}\Phi,
\]
for any vector $\Phi\in\HS_E$. 
Details are to be found in~\cite[Section~7]{FP02B}.

This treatment of `continuous measurement' is certainly
the coarsest possible. To examine more deeply the relationship
between the two manifestations of the Zeno effect, one should
consider more refined models for the interaction of a quantum
with a classical system\eg as in~\cite{BJ93}.
\section{Equilibrium States for Zeno Dynamics}
\labelS{equilibrium}
The explicit form of the generator of the Zeno dynamics found in~\refP{reduced} 
readily provides us with examples for equilibrium
states for the Zeno dynamics: Every equilibrium state for the
reduced dynamics $U_E$ will be one, since the original representation
of $\OA_E$ on $\HS_E$ is faithful and thus the automorphism groups
$\tau^E$ and $\widehat{\tau}^E$ of $\OA_E$ are identical:
\begin{Lcoro}{explicit-KMS}
  If $(U,E)$ is regular and satisfies~\eqref{eq:AZC} for \OA, then, for every $\beta>0$, 
the set of $(\tau^E,\beta)$-KMS states of $\OA_E$ equals
the set of $(\widehat{\tau}^E,\beta)$-KMS states.
\end{Lcoro}
This result is independent of the representation, 
since the reasoning of \refP{reduced}
can be repeated in any covariant representation.
It applies, in particular,
to the important case of Gibbs states as we will now show 
for quantum spin systems. 
The ordinary local Gibbs states over bounded regions $\Lambda$ are
\[
\omega_\Lambda(A)\DEF
\frac{\Tr_{\HS_\Lambda} 
\bigl(\ee^{-\beta H(\Lambda)}A\bigr)}{%
\Tr_{\HS_\Lambda}\bigl(\ee^{-\beta H(\Lambda)}\bigr)},
\quad\text{for } A\in\OA(\Lambda),
\]
and a candidate for a local Zeno equilibrium over $\Lambda$ is thus
\[
\omega_{E_\Lambda}(A_{E_\Lambda})\DEF
\frac{\Tr_{\HS_{\Lambda}} 
\bigl(\ee^{-\beta E_\Lambda H(\Lambda) E_\Lambda}A_{E_\Lambda}\bigr)}{%
\Tr_{\HS_{\Lambda}}\bigl(\ee^{-\beta E_\Lambda H(\Lambda) E_\Lambda}\bigr)},
\quad\text{for } A_{E_\Lambda}\in\OA(\Lambda)_{E_\Lambda},
\]
if $E_\Lambda\in\OA(\Lambda)$ is some collection of projections, and
where, as before, $\OA(\Lambda)_{E_\Lambda}=E_\Lambda \OA(\Lambda) E_\Lambda$.
Here it is safe to take the trace over the full local space $\HS_\Lambda$, since 
$\omega_{E_\Lambda}(A B_{E_\Lambda} C)=
\omega_{E_\Lambda}(A_{E_\Lambda} B_{E_\Lambda} C_{E_\Lambda})$,
for $A$, $B$, $C\in\OA(\Lambda)$, as follows easily from
$E\ee^{-\beta E_\Lambda H(\Lambda)E_\Lambda}=
\ee^{-\beta E_\Lambda H(\Lambda)E_\Lambda}E=
\ee^{-\beta E_\Lambda H(\Lambda)E_\Lambda}$ 
and the invariance of the trace under cyclic permutations.

Assume that the local dynamics $\tau_t^\Lambda$ generated by $H(\Lambda)$
converges uniformly  to an automorphism group $\tau$ of \OA. Then
we know~\cite[Proposition~6.2.15]{b:BR79/81}, that every thermodynamic
limit point of the ordinary local Gibbs states,
that is, a weak* limit of a net of extensions $\omega_\Lambda^G$ of 
$\omega_\Lambda$ to \OA, is a $(\tau,\beta)$-KMS state over \OA.
As a direct consequence of these considerations 
and~\refC{explicit-KMS}, 
we obtain those equilibrium states for the Zeno dynamics which are limits
of local Gibbs states.
\begin{Lcoro}{gloGibbs}
Let $\beta>0$. Let $\Lambda_\alpha\to\infty$ be such that the local dynamics
converges uniformly to the global dynamics $\tau$, 
and the net of local Gibbs states $\omega_{\Lambda_\alpha}$ 
has a thermodynamic limit point $\omega$. 
Let $U$ be the unitary group representing $\tau$ 
in the GNS representation of $\omega$.
If a sequence of projections $E_{\Lambda_\alpha}\in\OA(\Lambda_\alpha)$
converges in norm to a projection $E$ in \OA
such that $(U,E)$ is regular and satisfies~(AZC),
then $\omega_E(A_E)\DEF\lim_\alpha\omega_{E_{\Lambda_\alpha}}^G(A_E)$ 
defines  a $(\tau^E,\beta)$-KMS state on $\OA_E$.
\end{Lcoro}
\begin{proof}
  The local Gibbs states $\omega_{E_{\Lambda_\alpha}}$ are the unique $\beta$-KMS states
on the finite-dimensional algebras $\OA_{\Lambda_\alpha}$ for the 
reduced dynamics $\widehat{\tau}^{E_{\Lambda_\alpha}}$.
If $\{E_{\Lambda_\alpha}\}$ converges uniformly, these
local Gibbs state possess $\omega_E$ as a weak-* limit,
which is a KMS state on $\OA_E$
for the reduced dynamics $\widehat{\tau}^E$
associated with $\tau$.
Then, by ~\refC{explicit-KMS}, $\omega_E$ is also a $(\tau^E,\beta)$-KMS state.
\end{proof}
\begin{nxmpl}\label{ex:qspin-equi}
Let us review Example~\ref{ex:qspin}, and assume again that the interaction
$\Phi$ is such that the global surface energy $W_\Phi(\Lambda)$
is bounded. Then, given the family of projections $E_{\varphi_\Lambda;\Lambda'}$ 
of Example~\ref{ex:qspin}, and $E_{\varphi_\Lambda}$ being its uniform 
limit point, $(U,E_{\varphi_\Lambda})$ satisfies~AZC, 
and is regular as discussed in \refS{EHE}. 
Thus the conditions of \refC{gloGibbs} are satisfied, and 
we have to look at the state
\[
\omega_{E_{\varphi_\Lambda}}(A_{E_{\varphi_\Lambda}})\DEF
\lim_{\Lambda'\to\infty}
\frac{
  \Tr_{\HS_{\Lambda'}} 
      \bigl(\exp(-\beta\, 
                   \bbone_{\Lambda'\setminus\Lambda}\otimes P_{\Phi_{\Lambda}}
                   \, H(\Lambda') \,
                    \bbone_{\Lambda'\setminus\Lambda}\otimes  P_{\Phi_{\Lambda}})
A_{E_{\varphi_\Lambda;\Lambda'}}\bigr)
     }{\Tr_{\HS_{\Lambda'}}  
          \bigl(\exp(-\beta\,  \bbone_{\Lambda'\setminus\Lambda}\otimes  
                  P_{\Phi_{\Lambda}}
                  \, H(\Lambda') \,
\bbone_{\Lambda'\setminus\Lambda}\otimes  P_{\Phi_{\Lambda}})\bigr)
  }
\]
where $A_{E_{\varphi_\Lambda;\Lambda'}}\in 
\OA_{E_{\varphi_\Lambda;\Lambda'}}$ converges in $\OA_{E_{\varphi_\Lambda}}$ to
$A_{E_{\varphi_\Lambda}}$. This limit defines
a $(\tau^{E_{\varphi_\Lambda}},\beta)$-KMS state on $\OA_{E_{\varphi_\Lambda}}$.
If, in the decomposition $\OA=\OA_{\Lambda^c}\otimes\OA_\Lambda$, the global
Hamiltonian decomposes as 
\[
H=\sum_i H_{\Lambda^c,i}\otimes H_{\Lambda,i}
\]
then $\omega_{E_{\varphi_\Lambda}}$ is exactly the
Gibbs equilibrium for the \textbf{averaged Hamiltonian}
\[
\EE^{\varphi_\Lambda}(H)=
\sum_i  H_{\Lambda^c,i} \cdot\varphi_{\Lambda}(H_{\Lambda,i})
\]
with respect to the local state $\varphi_{\Lambda}$
over the interior region $\Lambda$.
This state has been shown in~\cite[Section~III]{FW95} to be the
strong-coupling limit of equilibrium states for the Hamiltonians
$H_\lambda=H+\lambda \bbone_{\Lambda^c}\otimes P_{\Phi_{\Lambda}}$,
in accordance with our results in the last section. Thus the Zeno 
dynamics effectively decouples the the interior $\Lambda$ from
the exterior part $\Lambda^c$ of the system, while the influence of the
interior is reduced to a `mean field'-type interaction, 
where the inner part of the system is averaged 
out with respect to the chosen local state $\varphi_\Lambda$.
\end{nxmpl}
The above special result for Gibbs states has a counterpart for
states which satisfy a \emph{maximum entropy
condition}: Let $\omega$ be a faithful, normal state on the 
von Neumann algebra \OA. For the definition of the \textbf{relative
entropy} $S(\omega,\varphi)$ of a state $\varphi$ on \OA with respect to 
$\omega$ we refer the reader to~\cite[Definition~6.2.29]{b:BR79/81}.
Raggio and Werner have shown the following general result:
\begin{theo*}[{\cite{RW90}}, Theorem~7]
  Let $\widetilde{\omega}$ be a state on \OA with 
$\widetilde{\omega}(E)=1$. Then holds  the estimate
$ S(\omega,\widetilde{\omega})\geq -\log\bigl(\omega(E)\bigr)$,
with equality if and only if $\omega\bigl([E,A]\bigr)=0$,
and $\widetilde{\omega}(A)=\omega(EAE)/\omega(E)$,
for all $A\in\OA$.
\end{theo*}
The state $\widetilde{\omega}$ with $\widetilde{\omega}(E)=1$
is a natural candidate for a Zeno equilibrium state on $\OA_E$.
If the original state $\omega$ is a Gibbs state and the Hamiltonian
commutes with $E$, then this conforms with our above result. 
However, these restrictions are too severe to identify
general Zeno equilibria, which will therefore in general not
maximise the relative entropy on the total
algebra \OA.

As a final application of our theoretical results, we reconsider
the model of Example~\ref{ex:XY}. As noted there, the 
$\tau^E$-invariant state we chose was not an equilibrium state.
We are now in a position to correct this.
\begin{nxmpl}[Zeno Equilibria in the $X$-$Y$ Model]
\label{ex:XYZ}
  We start by choosing again a fixed state $\rho_0\in\HS_0$ over
the centre site of the chain. Again, we use
\[
E_{\rho_0}\DEF \bbone_{\HS_{[-\infty,-1]}}
               \otimes P_{\rho_0}\otimes
               \bbone_{\HS_{[1,\infty]}}
\]
as the Zeno projection.
Since in this model the interaction is has range $1$, and the projection
acts local, the Zeno dynamics $\tau^{E_{\rho_0}}$
will certainly exist, by the reasoning of Example~\ref{ex:qspin}.
As in~Example~\ref{ex:qspin-equi}, the local Zeno Hamiltonians
decompose into two commuting, nontrivial parts over the 
subchains $[-m,-1]$ and $[1,n]$ 
which are averaged with respect to $\rho_0$, and a scalar part:
\[
E_{\rho_0}HE_{\rho_0} = 
H_{-m}^{\rho_0} + H_0^{\rho_0} + H_{+n}^{\rho_0},
\]
where $E_{\rho_0}$ is restricted to $\HS_{[-m,n]}$ in the natural way.
Explicitly we obtain
$
H_0^{\rho_0}=h\rho_0(a_0^\ast a_0)
$, and 
\[
H_{+n}^{\rho_0}=
\frac{J}{2}\bigl(\overline{\rho_0(a_0)}a_1+\rho_0(a_0)a_1^\ast\bigr)+
H_{[1,m]},
\]
and likewise for $H_{-m}^{\rho_0}$. Straightforwardly, we obtain
Gibbs states over the left and right subchains:
\[
\omega_{\rho_0,\beta}^+(A_+)\DEF\lim_{n\to\infty}
\frac{\Tr_{\HS_{[1,n]}}\bigl(\ee^{-\beta H_{+n}^{\rho_0}} A_{+n}\bigr)}{%
\Tr_{\HS_{[1,n]}}\bigl(\ee^{-\beta H_{+n}^{\rho_0}}\bigr)},
\]
where 
$\{A_{+n}\in\OA_{[1,n]}\}$ converges to $A_+$ in
$\OA_{[1,\infty]}$, which is the weak closure of the union of the local algebras
$\OA_{[1,n]}$.
A similar result holds  for the Gibbs state 
$\omega_{\rho_0,\beta}^-$
over the left subchain. 
Now the observables for the Zeno dynamics are in this model
all of the form
\[
A_{E_{\rho_0}}=\sum_i\rho_0(A_{0,i})A_{-,i}
\otimes P_{\rho_0} \otimes A_{+,i}
\]
where $A_{±,i}\in\OA_±$, $A_0\in\OA_0$, since the local observables
are nothing but polynomials in the local generators $a_x$, $a_x^\ast$. 
Thus, since the scalar factor
$\ee^{-\beta H_0^{\rho_0}}$ cancels out in the definition of the
Gibbs state, we finally obtain the global equilibrium state
on $\OA_{E_{\rho_0}}$:
\[
\omega_{\rho_0,\beta}(A_{E_{\rho_0}})\DEF
\sum_i
\rho_0(A_{0,i})
\omega_{\rho_0,\beta}^-(A_{-,i})
\omega_{\rho_0,\beta}^+(A_{+,i}),
\]
or 
\[
\omega_{\rho_0,\beta}=\omega_{\rho_0,\beta}^-
\otimes \rho_0 \otimes
\omega_{\rho_0,\beta}^+.
\]
This is the desired equilibrium state for the Zeno dynamics 
$\tau^{E_{\rho_0}}$. Moreover, it is the unique
$(\tau^{E_{\rho_0}},\beta)$-KMS state on $\OA_{E_{\rho_0}}$,
since the Gibbs states are the only KMS states in this
class of models, as is shown in~\cite[Appendix]{ROB73}
for the original spin chain by a calculation which depends
only on the local CAR, and therefore also applies to
the Zeno dynamics (this can also be seen by more general
arguments~\cite[Theorem~6.2.47]{b:BR79/81}).
The state $\omega_{\rho_0,\beta}$ decomposes into a product state with respect
to the decomposition $\{[-\infty,-1],0,[1,\infty]\}$ of the chain,
which again shows that the Zeno dynamics decouples the left and right
subchains. The equilibria of the lateral subchains are determined by 
$\rho_0$-averaged Hamiltonians, imposing boundary conditions as
already exhibited in Example~\ref{ex:qspin-equi}. For varying $\rho_0$, 
these equilibria are parametrised by
$\rho_0(a_0)$.
Yet there is somewhat more to say about this example: For here, the
difference between the Zeno Hamiltonian and the original one
is a finite combination of local generators 
$a_x$, $a^\ast_x$, $x=0,\ ±1$,  
as can easily be seen from the explicit forms of $H$ and 
$E_{\rho_0}HE_{\rho_0}$. This difference is therefore
a bounded operator, and moreover it is entire analytic
for $\tau^{E_{\rho_0}}$. Thus  the original dynamics
is a local  perturbation of the Zeno dynamics.
Under these conditions, the
general results about the return to equilibrium~\cite[Theorem~2]{ROB73}
imply that the system starting in a global equilibrium state
for the dynamics defined by $H$ will eventually evolve
toward a KMS state for the Zeno dynamics. In fact it will approach
the Zeno Gibbs state constructed above, since it
is the unique KMS state as seen before.
\end{nxmpl}
The property of the Zeno dynamics to spontaneously 
approach a $(\tau^E,\beta)$-KMS 
state does only depend on the properties of $EHE-H$. We conclude
our discussion by noting this fact:
\begin{Lcoro}{Zeno-equi}
Let $(U,E)$ be regular and satisfy~\eqref{eq:AZC} for \OA.
  Let $\omega|_{\OA_E}$ be the restriction of a $(\tau,\beta)$-KMS state
of \OA to $\OA_E$. Assume that $(\OA_E,\tau^E)$ is asymptotically
Abelian, and that $H-EHE$ is entire analytic for $\tau^E$.
Then, every weak* limit point for $t\to±\infty$ 
of $\tau^E_t\omega|_{\OA_E}$ 
is a $(\tau^E,\beta)$-KMS state.
\end{Lcoro}
\section{Conclusions}\labelS{conclusions}
The present and our previous work~\cite{AUS02A} have demonstrated
that quantum statistical mechanics is another natural field for
the exploration of the Zeno effect. In view of our general
estimation of the status of the effect (see below), we think 
the examples shown to be the most important part of 
our work. 
Let us review the final Example~\ref{ex:XYZ}:
The decomposition of the global Gibbs equilibrium of the $X$-$Y$
model into a product state under the special Zeno dynamics is 
hardly surprising. In fact, this behaviour is characteristic for 
Gibbs states, when the boundary interaction is
removed~\cite[Definition~6.2.16]{b:BR79/81}.
But the new, and physically remarkable point in Example~\ref{ex:XYZ}
is that a frequent observation at the microscopic
level, even a single site, leads to a different equilibrium,
namely that of two isolated subchains with a boundary condition.
Moreover, we have seen that this behaviour is dynamically 
observable in the sense that the chain prepared in the Gibbs 
equilibrium will evolve to the lateral Zeno equilibria, under the
Zeno dynamics. This shows that the context
of quantum statistical mechanics can indeed exhibit 
new phenomenological aspects of the quantum Zeno effect.
To give a tenable outlook toward further developments,
it seems 
appropriate to give some epistemological rationale as to
why the Zeno effect is worth any consideration at all.
For its theoretical explanatory power is very limited, due to
the very reason which lends it its heuristic appeal: It is
an extremely generic phenomenon. But its ubiquity
renders its value for basing theoretical explanations for 
physical phenomena on it small. 
The effect therefore seems more interesting if considered in
special model cases, where it can yield real, and sometimes
surprising, predictions of phenomena.
Quantum statistical mechanics might provide a fruitful
ground in that respect:
Since characteristic lifetimes
are generally longer for collective than for single- or 
few-particle phenomena, it is conceivable that the 
Zeno effect is easier to detect in this context than in 
many experiments devised so far in
atomic and particle physics, see the reviews 
in~\cite{WHI00,NNP96}. 
To give a theoretical treatment of further interesting 
phenomena,  one would need 
a truly representation-independent formulation of the results
shown here and in~\cite{AUS02A}, as well as an independent
treatment of the $C^\ast$-case. This is work in progress.
\newcommand{\noopsort}[1]{} \newcommand{\singleletter}[1]{#1}
  \providecommand{\noopsort}[1]{} \providecommand{\singleletter}[1]{#1}
\providecommand{\bysame}{\leavevmode\hbox to3em{\hrulefill}\thinspace}
\providecommand{\MR}{Math.~Rev.\ }
\end{document}